\documentclass[aps,prb,twocolumn,superscriptaddress,showpacs,preprintnumbers]{revtex4-2}
\usepackage{amsmath,amssymb,bm}
\usepackage{graphicx}
\usepackage{hyperref}
\usepackage{float}
\usepackage{comment}

\begin{document}

\title{Optical spectroscopy of composite fermion edge states in the fractional quantum Hall effect}

\author{Maria Sebastian}
\email{mariasebastian@tamu.edu}
\affiliation{Department of Physics and Astronomy,
Texas A\&M University, College Station, Texas 77843, USA}

\author{Ashutosh Singh}
\email{asingh.n19@gmail.com}
\thanks{Current address:
        School of Physics and Optoelectronic Engineering, Hainan University, Haikou 570228, China}
\affiliation{Department of Physics and Astronomy,
Texas A\&M University, College Station, Texas 77843, USA}

\author{Alexey Belyanin}
\email{belyanin@tamu.edu}
\affiliation{Department of Physics and Astronomy,
Texas A\&M University, College Station, Texas 77843, USA}

\date{\today}

\begin{abstract}

We show that edge states in fractional quantum Hall effect samples can be selectively probed and excited with  sub-terahertz optical spectroscopy.
Using the composite fermion (CF) mean-field framework, which maps the
strongly correlated fractional quantum Hall problem
onto an effective integer quantum Hall problem, 
we calculate the absorbance spectrum for the Jain sequence of 
filling fractions including both bulk and edge states. The CF edge-state absorption
peaks appear in the millimeter-wave to sub-terahertz range, e.g., $60$--$500$\,GHz at $B = 10$\,T in GaAs, i.e. they are blueshifted with respect to the bulk CF cyclotron frequency but are well below the integer quantum Hall cyclotron frequency 
scale at the same magnetic fields. The number of resolved peaks in each series of the absorption spectrum counts the filled $\Lambda$-levels and fingerprints the fraction.
Inversion symmetry breaking near the  edge
activates optical transitions forbidden in the bulk 
and enables second-order nonlinear processes in electric-dipole approximation. 
The absolute frequency scale of the spectrum is set by the CF effective mass $m^*_{\rm CF}$, 
which is generated entirely by electron-electron interactions, so the absorption spectrum provides a direct optical probe of this interaction-induced mass.


\end{abstract}

\maketitle
\section{Introduction}
\label{sec:intro}

There is strong renewed interest in the studies of fractional Quantum Hall (FQH) effect, motivated by the possibility to realize these states both in real and synthetic systems and by the promise to utilize edge states for electron interferometry and topologically protected quantum computation \cite{Chamon, PhysRevB.73.245311,PhysRevLett.97.186803, PhysRevB.82.085321,PhysRevB.105.165310, Nakamura2020,Ron2021, greiner2023,nak23,mak2024,doi:10.1126/science.ado3912}. Although most of the studies dealt with DC response and transport measurements, several optical and microwave techniques have been 
proposed or implemented for spectroscopy or cavity-induced modification of of FQH states. Existing studies have focused
primarily on bulk collective
excitations~\cite{kukushkin2002cyclotron,dujovne2005composite,
rhone2011higher,faist2025} or collective bosonic edge plasmon
modes~\cite{france2025electrically,winter2025fractional},
rather than electronic states 
at the edge. While composite fermion (CF) edge-state transport theory has been
developed using single-particle $\Lambda$-level
wavefunctions~\cite{kirczenow1996composite,kirczenow1998composite},
the optical matrix elements and absorbance spectra of
these states have not been computed. The 
optical response of CF $\Lambda$-level edge states thus remains an open
problem, especially at 
the CF cyclotron frequency scale, which is of main interest in the characterization of these states. 

In this paper we present the theory of the optical response of 2D electron systems in the FQH regime, including both bulk and edge states, using the CF mean-field 
approach~\cite{jain1989composite,lopez1991fractional,halperin1993theory}. Within this approach, we were able to develop a semi-analytic, physically intuitive picture of the optical response, which allows one to clearly separate contributions of bulk and edge states and excite a particular edge state on demand. 

We compute the CF $\Lambda$-level eigenvalue structure, 
optical dipole matrix elements, and absorbance spectra
 for filling fractions in the Jain sequence $\nu = p/(2p+1)$, where $p = 1,2,\dots$. 
Similarly to the integer quantum Hall (IQH) case~\cite{singh2024coherent,singh2025valley},
optical transitions between CF $\Lambda$-level edge
states have significantly different transition energies
and polarization selection rules compared to the bulk,
enabling frequency- and polarization-selective
spectroscopy of individual CF edge channels without
disturbing the FQH bulk. The CF edge-state absorption peaks appear 
in the millimeter-wave to sub-terahertz range (e.g., 60-500 GHz at
$B=10$\,T in GaAs), well below the IQH cyclotron
frequency $\omega_c \approx 4$\,THz in the same
field, reflecting both the reduced effective field
$B^* = B/(2p+1)$ and the enhanced CF effective mass
$m^*_{\rm CF} \gg m^*_e$. The number of resolved peaks directly counts
the Jain index $p$ and identifies the filling fraction
$\nu$, providing a spectroscopic fingerprint
of the FQH state. Because the entire spectrum scales with
$\omega^*_c = eB^*/m^*_{\mathrm{CF}}c$ and $B^*$ is fixed by the
filling fraction, the absolute frequencies determine the CF effective
mass $m^*_{\mathrm{CF}}$~\cite{du1994drastic,villegas2022composite}.
This mass is generated entirely by electron-electron interactions and
its theoretical value remains uncertain~\cite{girlich1997}. Furthermore,
inversion symmetry breaking near the edge
activates $\Delta p = \pm 2$ transitions that are forbidden in the bulk 
and generates nonzero permanent dipole differences
between consecutive $\Lambda$-levels, thus allowing second-order nonlinear processes in the electric-dipole approximation.

\section{Composite Fermion Edge States}
\label{sec:eigen}

 At filling fraction $\nu = 1/3$ the lowest Landau
level is one-third filled and the electrons are strongly interacting.
Attaching two magnetic flux quanta $\phi_0 = hc/e$ to each electron
produces a new quasiparticle, the composite fermion, which sees a
reduced effective magnetic field and behaves much more like a free
particle under integer quantum Hall effect
conditions~\cite{jain1989composite}. Quantitatively, the original
electrons see a field $B$ with electron density $n_e = \nu B/\phi_0$.
Attaching two flux quanta subtracts $2\phi_0 n_e$ from the field the CF
experiences, leaving the residual effective field:
\begin{equation}
B^* = B - 2\,\phi_0\,n_e = B(1-2\nu).
\label{eq:Bstar_nu}
\end{equation}
For the Jain sequence $\nu = p/(2p+1)$, substituting gives:
\begin{equation}
B^* = \frac{B}{2p+1},
\label{eq:Bstar_jain}
\end{equation}
so for $\nu = 1/3$ ($p=1$):
\begin{equation}
B^* = \frac{B}{3}.
\label{eq:Bstar_13}
\end{equation}
The CF now fill $p$ integer $\Lambda$-levels in $B^*$, just like free
electrons fill $n$-labeled Landau levels in the corresponding IQH problem, in which the
occupied levels are $n = 0$ through $n = p-1$. For $p = 1$
($\nu = 1/3$), only the $n = 0$ level is filled. The associated CF scales are
\begin{equation}
\ell^* = \sqrt{\frac{\hbar c}{eB^*}} = \ell_c\sqrt{3},
\qquad
\omega^*_c = \frac{eB^*}{m^*_{\mathrm{CF}}\,c},
\label{eq:CFparams}
\end{equation}
where $\ell_c = \sqrt{\hbar c/eB}$ is the electron magnetic
length. 

\medskip

In the Chern-Simons mean-field approximation~\cite{lopez1991fractional,halperin1993theory},
once the attached flux is replaced by its average
value, the CF Hamiltonian is:
\begin{equation}
H_{\mathrm{CF}} = \frac{1}{2m^*_{\mathrm{CF}}}
\!\left(\mathbf{p} + \frac{e}{c}\mathbf{A}^*\right)^{\!2}
+ V(y),
\label{eq:HCF}
\end{equation}
where $\mathbf{A}^*$ corresponds to $B^*$ and $V(y)$ is the
edge confining potential. This is formally identical to the
electron Hamiltonian solved in our previous work ~\cite{singh2024coherent}, with 
starred quantities replacing unstarred ones. 

We use the Landau gauge $\mathbf{A}^* = -yB^*\hat{x}$, with the
sample edge at $y=0$ and the physical sample occupying $y>0$.
Since the Hamiltonian has no explicit $x$-dependence, 
we write $\Psi(x,y)=e^{ikx}\chi(y)$, where $k$ is the conserved
$x$-component of momentum. Here we consider a hard-wall boundary condition $\chi(0)=0$ for simplicity. The
precise shape of the edge does not matter as long as it is sharper than
the magnetic length. Recent tunneling spectroscopy measurements of edge state dispersion agree well with the hard-wall boundary condition ~\cite{henok}. 

The equation for $\chi(y)$ is
\begin{equation}
-\frac{\hbar^2}{2m^*_{\mathrm{CF}}}
\frac{\partial^2\chi}{\partial y^2}
+ \frac{1}{2}m^*_{\mathrm{CF}}\omega^{*2}_c
(y - y^*_k)^2\,\chi = E_{nk}\,\chi,
\label{eq:schrodinger_full}
\end{equation}
where $y^*_k = k\ell^{*2}$ is the center of the CF cyclotron
orbit. Introducing dimensionless variables
$\tilde{y}=y/\ell^*$ and $\tilde{E}_{nk}=E_{nk}/\hbar\omega^*_c$,
and using $m^*_{\mathrm{CF}}\omega^*_c/\hbar = 1/\ell^{*2}$,
Eq.~(\ref{eq:schrodinger_full}) reduces to 
\begin{equation}
\frac{\partial^2\chi}{\partial\tilde{y}^2}
+ \left[2\tilde{E}_{nk}
- \left(\tilde{y} - k\ell^*\right)^2\right]\chi = 0,
\quad \chi(\tilde{y}\to\infty)=0.
\label{eq:dimensionless}
\end{equation}

\begin{figure}[tb]
\centering
\includegraphics[width=\columnwidth]{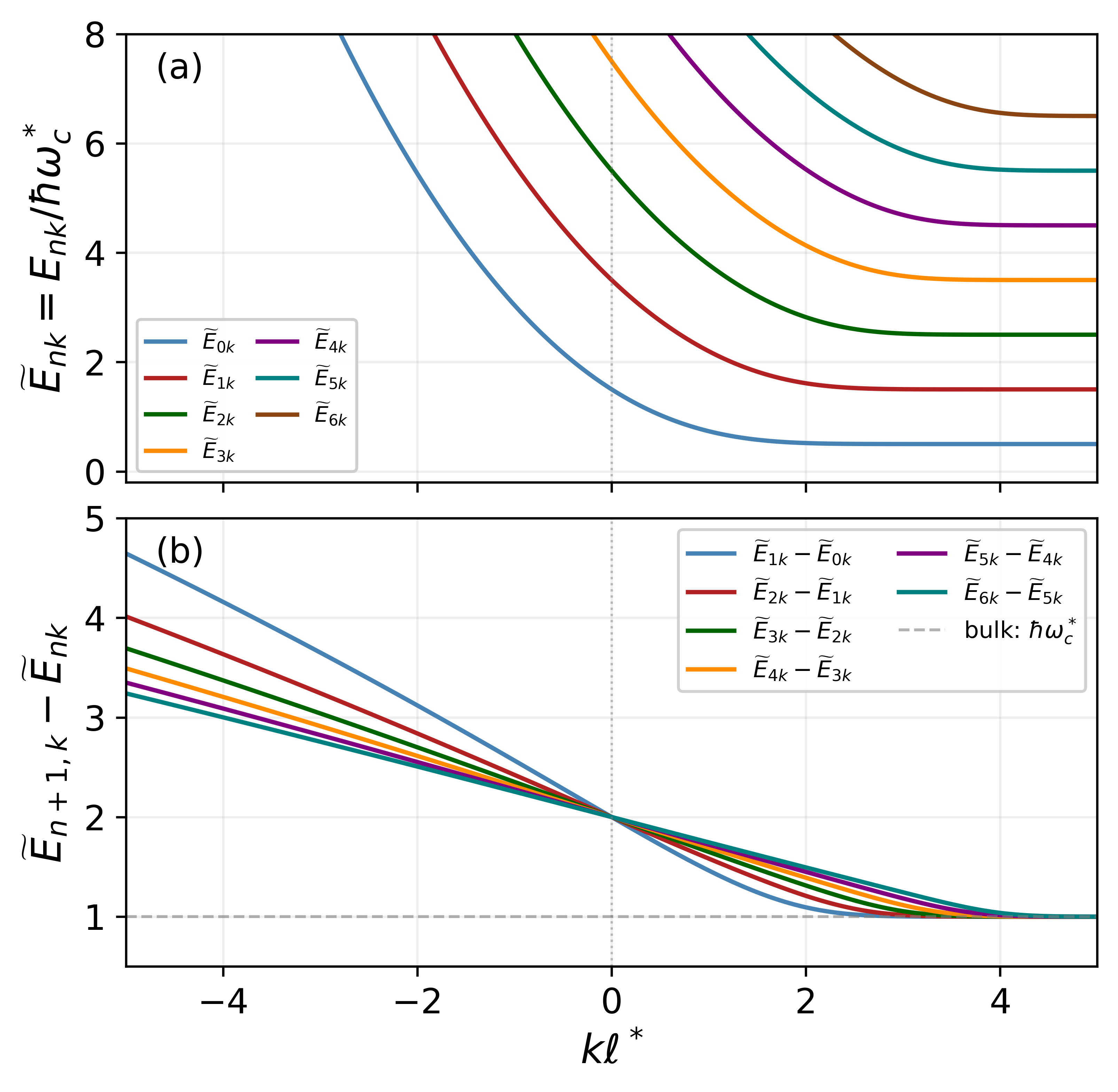}
\caption{CF $\Lambda$-level dispersion near a hard-wall edge.
All quantities are dimensionless. Energies are normalized by
$\hbar\omega^*_c$ and momentum by $\ell^*$. (a) The seven lowest
$\Lambda$-level energies $\widetilde{E}_{nk} = E_{nk}/\hbar\omega^*_c$
as a function of $k\ell^*$, with $n = p-1$. Levels are flat and equidistant in the
bulk (large positive $k\ell^*$) and bend upward nonuniformly near
the edge (negative $k\ell^*$). 
 (b) Consecutive transition energies
$\widetilde{E}_{n+1,k} - \widetilde{E}_{nk}$ for $n = 0$ to $5$. All approach the bulk cyclotron
energy $\hbar\omega^*_c$ (dashed line) for large positive $k\ell^*$
and rise to higher frequencies near the edge.}
\label{fig:dispersion}
\end{figure}

We solve Eq.~(\ref{eq:dimensionless}) numerically and calculate the dipole matrix elements of the optical transitions and the resulting absorbance spectrum. 
The energy eigenvalues are shown in Fig.~\ref{fig:dispersion}.
Panel (a) shows the seven lowest CF $\Lambda$-levels as a
function of $k\ell^*$. For large positive $k\ell^*$, the center of the cyclotron orbit is far from the edge and the states are bulk-like. The energy levels are equidistant and dispersionless, consistent with the bulk harmonic oscillator result $E^{\mathrm{bulk}}_{nk}
= \hbar\omega^*_c\!\left(n + \tfrac{1}{2}\right)$,
  $n = 0,1,2,\ldots$. As $k\ell^*$ decreases toward zero and then goes negative, the center of the cyclotron orbit crosses the edge, and the solution is completely dominated by the presence of the boundary. All energy levels disperse upward.Their spacing becomes nonuniform and all transition energies become higher than the transition energy $\hbar\omega^*_c$ between the bulk states. This provides spectral selectivity for edge-state spectroscopy and enables selective control of individual edge states.

Panel (b) shows six consecutive transition energies
$\widetilde{E}_{n+1,k} - \widetilde{E}_{nk}$ for
$n = 0$ to $5$. All six approach $\hbar\omega^*_c$ in
the bulk and rise above it near the edge.

\section{Dipole transition matrix elements}
\label{sec:matelem}
\begin{figure}[t]
\centering
\includegraphics[width=\columnwidth]{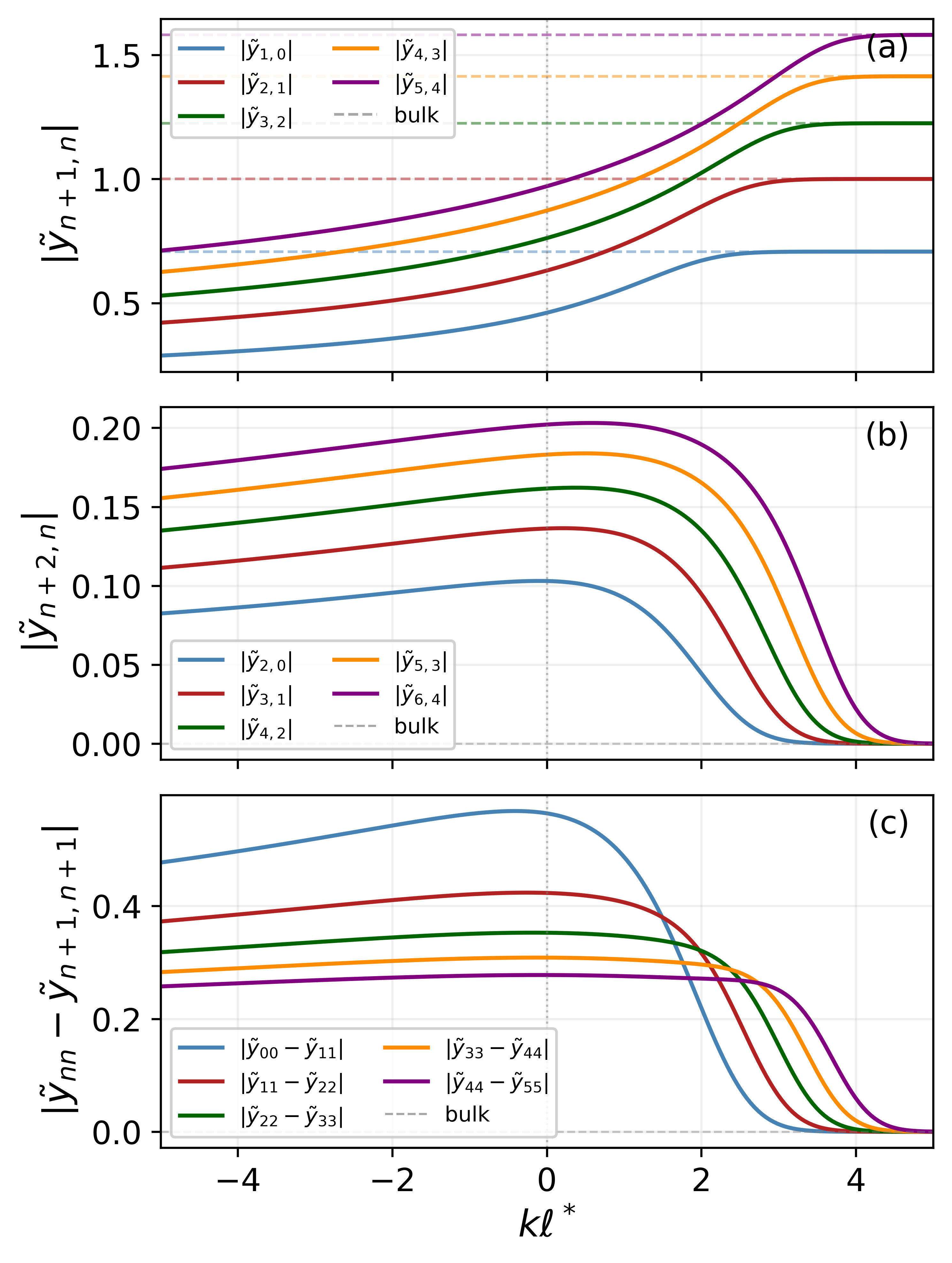}
\caption{CF dipole matrix elements as a function of the dimensionless
cyclotron orbit center $k\ell^*$. These are properties of the CF
$\Lambda$-levels, computed from the dimensionless eigenfunctions, and
are therefore independent of the filling fraction. All quantities are
dimensionless, normalized by $\ell^*$. (a) Matrix elements
$|\tilde{y}_{n+1,n,k}|$ of the transitions dipole-allowed in the bulk,
for $n = 0$ to $4$. Dashed lines show the bulk harmonic oscillator
values $\sqrt{(n+1)/2}$. Near the edge all matrix elements decrease
below their bulk values. (b) Transition matrix elements
$|\tilde{y}_{n+2,n,k}|$, forbidden in the bulk but nonzero near the
edge due to the breaking of inversion symmetry, for $n = 0$ to $4$.
(c) Permanent dipole differences
$|\tilde{y}_{nn,k} - \tilde{y}_{n+1,n+1,k}|$, zero in the bulk but
growing near the edge, for $n = 0$ to $4$.}
\label{fig:matelem}
\end{figure}

Next, we calculate the optical dipole matrix elements of the transitions
between $\Lambda$-levels, which determine the selection rules and the
absorbance spectra. Following~\cite{singh2024coherent}, the optical
response is governed by the dimensionless position matrix elements
\begin{equation}
\tilde{y}_{n'n,k}
= \frac{1}{\ell^*}\langle n'k\,|\,y\,|\,nk\rangle
= \int_0^\infty \chi_{n'k}(\tilde{y})\,
\tilde{y}\,\chi_{nk}(\tilde{y})\,d\tilde{y},
\label{eq:matelem}
\end{equation}
where $\chi_{nk}$ are the normalized eigenfunctions of
Eq.~(\ref{eq:dimensionless}). The current density matrix elements
follow from these through the transition frequency $\omega_{n'n,k}$.
The results are shown in Fig.~\ref{fig:matelem}. Panel (a) shows the
matrix elements of the transitions dipole-allowed in the bulk,
$|\tilde{y}_{n+1,n,k}|$ for $n = 0$ to $4$. Far from the edge (large
positive $k\ell^*$), they match their harmonic oscillator values
$\sqrt{(n+1)/2}$ (dashed reference lines), confirming the numerical
accuracy. Near the edge all decrease below their bulk values. Panels
(b) and (c) show the matrix elements for the $\Delta n = \pm 2$
transitions and the permanent dipole differences
$|\tilde{y}_{nn,k} - \tilde{y}_{n+1,n+1,k}|$, both of which vanish in
the bulk but become nonzero near the edge due to inversion symmetry
breaking. These terms give rise to the dipole-allowed second-order
optical nonlinearity, which can, for example, drive the optical
rectification quasi-DC photocurrent~\cite{singh2024coherent}.

\section{Absorbance Spectrum}
\label{sec:absorbance}
We now compute the dimensionless 2D absorbance spectrum. Light couples to the physical electron current, not to the composite
fermions directly. Flux attachment leaves the charge density and the
current operator unchanged, so the dipole coupling is identical to the
IQH case~\cite{singh2024coherent}, and near the edge the electron
current is organized into CF $\Lambda$-level channels. 
 Following the framework
of our IQH papers~\cite{singh2024coherent,singh2025valley}, the
$\hat{y}$-polarized absorbance is
\begin{align}
& \mathcal{A}_{\hat{y}}(\omega)
\propto \sum_{n,n',k}
\omega_{n'n,k}^2\,|\tilde{y}_{n'n,k}|^2\,
\bigl(F_{nk} - F_{n'k}\bigr) \nonumber \\
& \times \frac{\gamma/\pi}{\gamma^2 + (\omega - \omega_{n'n,k})^2},
\label{eq:absorbance}
\end{align}
where $\omega_{n'n,k} = (E_{n'k}-E_{nk})/\hbar$ is the transition
frequency, $F_{nk}$ is the occupation number, which we take as a Fermi
distribution at zero temperature for simplicity, and $\gamma$ is the
phenomenological broadening parameter.The overall normalization by the geometric overlap factor $f$ is
discussed below. The $\hat{x}$-polarized
absorbance $\mathcal{A}_{\hat{x}}$ is obtained by replacing
$\omega_{n'n,k}^2$ with $\omega_c^{*2}$ in
Eq.~(\ref{eq:absorbance}). Near the edge
$\omega_{n'n,k} \neq \omega^*_c$, and the difference
$\mathcal{A}_{\hat{y}} - \mathcal{A}_{\hat{x}}$ isolates the
contribution from edge states, canceling the dominant bulk cyclotron
resonance background~\cite{singh2024coherent,singh2025valley}.

\begin{figure}[t]
\centering
\includegraphics[width=\columnwidth]{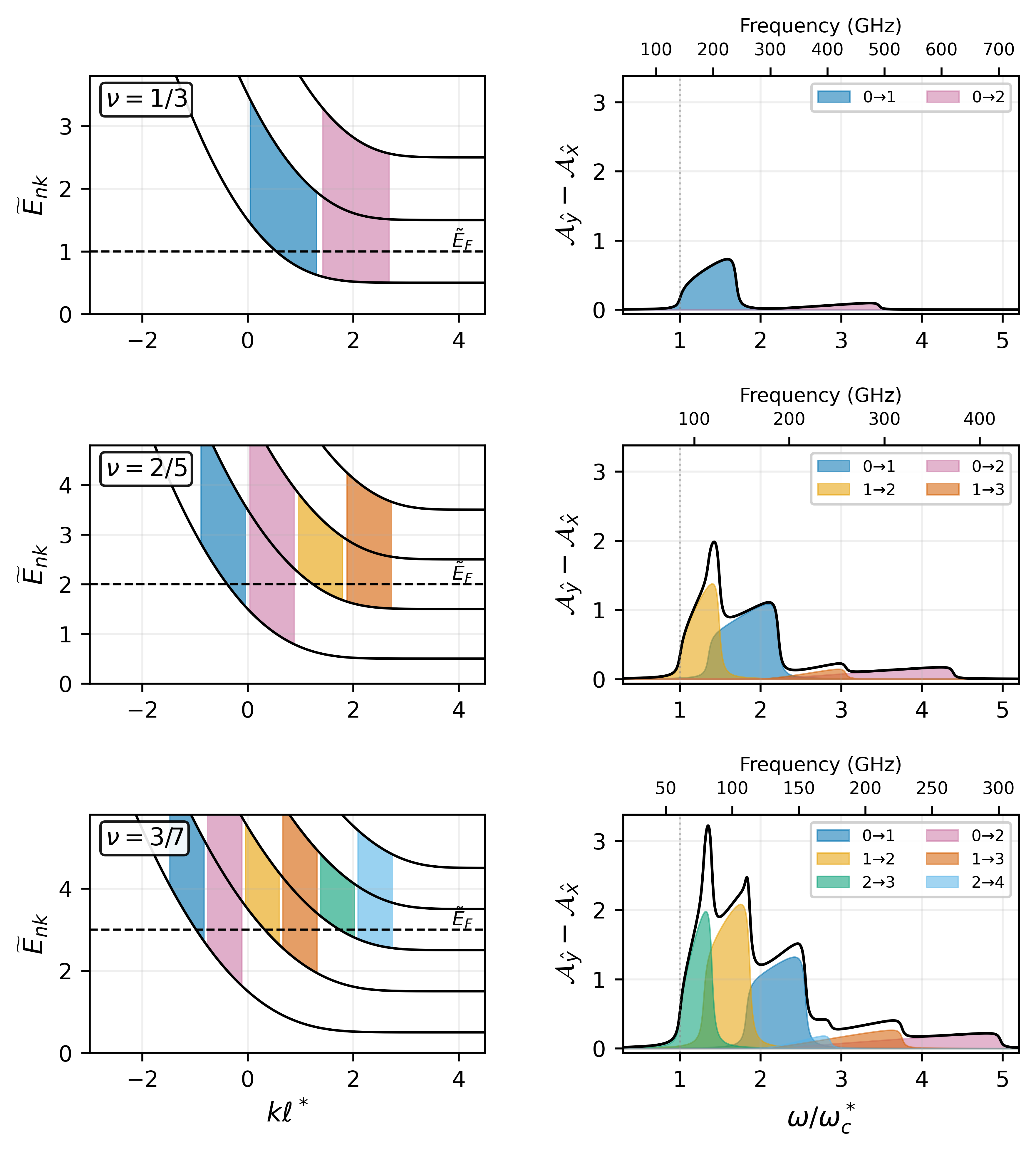}
\caption{Optical response of CFs for three filling fractions $\nu = 1/3$, $2/5$, and
$3/7$ (three positions of the CF Fermi level).   Left column:
$\Lambda$-level electron dispersion near the edge, with dashed line marking the normalized CF Fermi level $\widetilde{E}_F = p = 1,2,3$ from top to bottom. Shaded regions show the
phase-space windows that contribute to each transition, colored to
match the corresponding features in the spectra. Right column: dimensionless 
absorbance difference 
$\mathcal{A}_{\hat{y}} - \mathcal{A}_{\hat{x}}$ normalized by the geometric overlap factor (see the text) as a function of 
normalized frequency $\omega/\omega^*_c$ (bottom axis) and
absolute frequency in GHz (top axis) for GaAs at $B = 10$\,T, assuming 
$m^*_{\mathrm{CF}} = 0.66\,m_0$~\cite{villegas2022composite}, $\gamma = 0.03\,\omega^*_c$, and zero temperature. Shaded
regions show contributions from individual transitions and the black
curve is their sum. The response contains two series: The transitions $n \to n+1$ that are dipole-allowed in the bulk and the transitions $n \to n+2$ at higher frequencies that are forbidden in the bulk and activated  by inversion symmetry breaking at the edge. Each series  contains $p$ resolved features, so
the peak count provides a direct spectroscopic fingerprint of the
filling fraction. All features appear above $\omega^*_c$ (vertical
dotted line), reflecting the edge blue shift.}
\label{fig:absorbance}
\end{figure}

The results for all three filling fractions are shown in
Fig.~\ref{fig:absorbance}. The shades of the absorption peaks are
color-matched to the regions in the electron energy dispersion diagrams that contribute to
them. The difference spectrum
$\mathcal{A}_{\hat{y}} - \mathcal{A}_{\hat{x}}$ contains two series of
features that are direct consequences of the CF edge-state structure.
The principal series arises from the dipole-allowed
$\Delta n = 1$ transitions $n \to n+1$. All its peaks appear above
$\omega^*_c$, reflecting the blue shift of the CF $\Lambda$-levels near
the edge, and the bulk cyclotron resonance at $\omega^*_c$ is
completely removed in the difference spectrum. The number of resolved
peaks in this series equals the number of filled $\Lambda$-levels $p$,
providing a direct spectroscopic count of $p$ and hence of the filling
fraction $\nu = p/(2p+1)$. Among these peaks, the $n=0\to1$ transition
appears at the highest frequency and $n=p-1\to p$ at the lowest, a
consequence of the nonequidistant bending of the $\Lambda$-levels near
the edge visible in Fig.~\ref{fig:dispersion}(b).

A second, weaker series arises from the edge-activated
$\Delta n = 2$ transitions $n \to n+2$, which are forbidden in the bulk
and switched on by the inversion symmetry breaking at the boundary.
This series lies above the principal one, starting near
$2.8\,\omega^*_c$, and also contains $p$ resolved features, each
carrying roughly $10\%$ of the principal peak intensity. For
$\nu = 1/3$ the single $0 \to 2$ feature appears at $3.34\,\omega^*_c$
with $13\%$ of the principal peak height. The two series are separated
in both frequency and intensity, so the peak count of each
independently fingerprints the fraction.

As $\nu$ increases from $1/3$, the entire spectrum shifts
to lower absolute frequencies, since
$\hbar\omega^*_c = \hbar\omega_c/(2p+1)$ decreases. To convert to
absolute frequencies we use the CF effective mass
$m^*_{\mathrm{CF}} = 0.66\,m_0$, where $m_0$ is the free electron mass,
measured by~\cite{villegas2022composite}  in a 30\,nm GaAs quantum well at
density $n_e \approx 1.1\times10^{11}$\,cm$^{-2}$. At $B = 10$\,T this
gives 
\begin{align}
\hbar\omega^*_c(\nu=1/3) &= 0.585\,\text{meV}
\quad (141.4\,\text{GHz}), \notag\\
\hbar\omega^*_c(\nu=2/5) &= 0.351\,\text{meV}
\quad (84.8\,\text{GHz}), 
\nonumber \\
\hbar\omega^*_c(\nu=3/7) &= 0.251\,\text{meV}
\quad (60.6\,\text{GHz}),
\nonumber
\end{align}
placing all edge features in the millimeter-wave to sub-terahertz
range ($60$--$500$\,GHz), accessible to current millimeter-wave
transmission and resonant cavity spectroscopy techniques. Because the CF effective mass is not determined within the mean-field
theory and depends on density and magnetic
field~\cite{villegas2022composite}, all results are presented
normalized to $\omega^*_c$ and converted to absolute frequencies using
the experimental value~\cite{villegas2022composite}. Conversely, since
the entire spectrum scales with $\omega^*_c = eB^*/m^*_{\mathrm{CF}}c$
and $B^*$ is fixed by the filling fraction, the absolute frequencies
determine $m^*_{\mathrm{CF}}$, a complementary optical route to the
activation-gap method used
previously~\cite{du1994drastic,park1998activation}.

The absorbance spectra in Fig.~\ref{fig:absorbance}(b) are normalized by a dimensionless beam overlap factor $f = \frac{L_x}{l_x} \frac{ \ell^*}{l_y}$, where $L_x \sim $ a few $\mu$m is the sample length in x-direction along the edge, $\ell^*$ is in tens of nm, and $l_x l_y$ is the beam cross-section  on the sample. Since the
free-space wavelengths are in the millimeter range 
across the fractions, a subwavelength metallic structure such as a nanoresonator, nanoslit, or nanoantenna is needed to concentrate the field onto the sample and bring $f$ close to one  \cite{chen2014squeezing, seo2009terahertz, kim2018terahertz, toma2015, aglieri}. 


\section{Discussion and Limitations}
\label{sec:corrections}

We calculated the optical response of FQH samples within the CF framework and showed the feasibility of selective probing and excitation of CF edge states without affecting the states in the bulk. This became possible because of the frequency and polarization selectivity of edge state absorption, which is a generic property of energy dispersion and inversion symmetry breaking near the edge. The situation is qualitatively similar in this regard to the IQH results of our previous
work~\cite{singh2024coherent,singh2025valley}, a direct consequence
of the Jain CF mapping~\cite{jain1989composite} which reduces the
strongly correlated FQH problem to an effective single-particle
IQH problem in the starred parameters $B^*$, $\ell^*$,
$\omega^*_c$. 

There are also important differences. 
The CF cyclotron frequency $\omega^*_c = \hbar e B^*/m^*_{\rm CF}$
is much smaller than the electron cyclotron frequency at the
same field, both because $B^* = B/(2p+1)$ is reduced and
because $m^*_{\rm CF} \gg m^*_e$. At $B = 10$\,T in GaAs,
$\hbar\omega_c \approx 17$\,meV while $\hbar\omega^*_c$
ranges from $0.59$\,meV at $\nu=1/3$ to $0.25$\,meV at
$\nu=3/7$, placing the FQH
edge peaks in the millimeter-wave to sub-terahertz range
($60$--$500$\,GHz). Moreover, $\omega^*_c$ is fraction-dependent, since both $B^*$  and $m^*_{\rm CF}$ depend on the fraction. This makes the spectral position of the peaks a sensitive diagnostics of CF parameters. The sensitivity to  $m^*_{\rm CF}$ is especially important, as the mass is generated entirely
by electron-electron interactions and cannot be obtained
from the mean-field theory itself~\cite{du1994drastic,villegas2022composite}.
The edge absorption spectrum could therefore serve as a
sensitive optical/microwave route to determining $m^*_{\rm CF}$.

The results presented above are obtained within the CF
mean-field approximation, in which the attached flux quanta
are replaced by their average and the CFs move independently
in the effective field $B^*$. This is the standard and
well-established starting point for CF
physics~\cite{jain1989composite,lopez1991fractional,halperin1993theory} which should 
capture the leading-order optical response.
However, it is important to discuss the corrections that go beyond this level
and the limitations of the approach.


\textit{Residual CF-CF interactions at the Hartree level:}
Including residual CF-CF interactions at the Hartree level modifies the CF energy spectrum and the electron density distribution at the edge~\cite{chklovskii1995structure}.  In future work, we plan to implement this correction
by adding a self-consistent Coulomb potential to the CF Hamiltonian and
quantifying the resulting shift in peak positions relative to the
mean-field results.


\textit{Correction to the current operator:}
Within the Chern-Simons formulation of the CF
theory~\cite{lopez1991fractional,halperin1993theory}, the current
operator acquires an additional contribution when a CF moves and drags
its attached flux. This term is neglected at the mean-field level. Estimating its effect on the edge-state matrix elements would require going beyond
mean-field, for example through exact diagonalization of small systems,
and is left to future work. 

\textit{Limitations and more complex states:}
The present calculation is restricted to the principal
Jain sequence $\nu = p/(2p+1)$ for $p = 1, 2, 3$.
Particle-hole conjugate fractions ($\nu = 2/3$, $3/5$,
$4/7$) have more complex edge structures with both
forward and backward moving modes, and even-denominator
states such as $\nu = 5/2$ are believed to require non-Abelian
wavefunctions beyond the CF mean-field picture.
At $\nu = 1/2$ the CF $\Lambda$-level picture breaks
down entirely as CFs form a compressible Fermi
sea~\cite{halperin1993theory} with no cyclotron gap. The
collapse of the spectrum toward zero frequency applies
only along the Jain sequence approaching half-filling.
Extension to the reverse Jain sequence $\nu = p/(2p-1)$,
where composite fermions experience a negative effective
field $B^* < 0$, and to the four-flux-quanta family
$\nu = p/(4p \pm 1)$ represents a natural future direction.

\section{Acknowledgements}
A.B. and M.S. were supported in part by the Keck Foundation (Award No.~CRM:0132347). 

\bibliography{references,Ref2}
\bibliographystyle{apsrev4-2}
\end{document}